\documentclass[a4paper,11pt]{article}
\usepackage{pos}

\usepackage{amsmath}

\usepackage{amssymb}
\usepackage{array}
\usepackage{calc}
\usepackage{longtable}
\usepackage{multirow,booktabs}
\usepackage{relsize}
\usepackage{pstricks}
\usepackage{graphicx}
\usepackage{xspace}
\usepackage{units}
\usepackage{placeins}

\def\beq{\begin{equation}}  
\def\eeq{\end{equation}}
\def\({\left(}
\def\){\right)}
\def\[{\left[}
\def\]{\right]}

\newcommand{\sherpa}{S\protect\scalebox{0.8}{HERPA}\xspace}

\newcommand{\amegic}{A\protect\scalebox{0.8}{MEGIC}\xspace}
\newcommand{\comix}{C\protect\scalebox{0.8}{OMIX}\xspace}

\newcommand{\Caesar}{C\protect\scalebox{0.8}{AESAR}\xspace}
\newcommand{\Centauro}{C\protect\scalebox{0.8}{ENTAURO}\xspace}
\newcommand{\fastjet}{F\protect\scalebox{0.8}{AST}J\protect\scalebox{0.8}{ET}\xspace}
\newcommand{\rivet}{R\protect\scalebox{0.8}{IVET}\xspace}

\newcommand{\powheg}{P\protect\scalebox{0.8}{OWHEG}\xspace}

\newcommand{\zcut}{\ensuremath{z_{\text{cut}}}}
\newcommand{\alphaS}{\alpha_\text{s}\xspace}

\newcommand{\NLO}{\text{NLO}\xspace}    
\newcommand{\NLL}{\text{NLL}\xspace}

\newcommand{\NLOpNLL}{\ensuremath{\NLO+\NLL}\xspace}
    
\newcommand{\NLOpNLLp}{\ensuremath{\NLOpNLL^\prime}\xspace}

\newcommand{\MCatNLO}{\text{\textsc{MC@NLO}}\xspace}

\newcommand\HERA{H\scalebox{0.8}{ERA}\xspace}

\renewcommand{\logo}{\relax}

\title{Precision calculations for groomed event shapes at \HERA}

\author[a]{Max Knobbe}
\author*[b]{Daniel Reichelt}
\author[a]{Steffen Schumann}
\author[a]{Leon Stöcker}

\affiliation[a]{Institut f\"ur Theoretische Physik, Georg-August-Universit\"at G\"ottingen, G\"ottingen, Germany}

\affiliation[b]{Institute for Particle Physics Phenomenology, Durham University, Durham DH1 3LE, UK}

\emailAdd{max.knobbe@uni-goettingen.de}
\emailAdd{daniel.reichelt@durham.ac.uk}
\emailAdd{steffen.schumann@phys.uni-goettingen.de}
\emailAdd{leon.stoecker@uni-goettingen.de}

\abstract{The possibility to reanalyse data taken by the \HERA experiments
  offers the chance to study modern QCD jet and event-shape observables in
  deep-inelastic scattering production. In this contribution we present resummed 
  and matched predictions for the groomed invariant-mass event shape in neutral-current 
  DIS including the effect of grooming the hadronic final state using the soft-drop
  technique. Non-perturbative corrections from hadronisation are taken into
  account through parton-to-hadron level transfer matrices extracted from
  dedicated Monte Carlo simulations with \sherpa, including uncertainties
  extracted from replica tunes to data from the HERA experiments.} 

\FullConference{31st International Workshop on Deep Inelastic Scattering (DIS2024)\\
 8–12 April 2024\\
Grenoble, France\\}

\begin{document}
\maketitle

\section{Introduction}
Studying deep-inelastic scattering (DIS) has historically been one of the
primary methods to discover and investigate strong-interaction phenomena. With
several future projects like the EIC but also proposed projects such as the LHeC
and FCC-eh machines, it is of considerable interest to challenge the theoretical
and experimental methods currently employed at the LHC with tasks necessary for
DIS \cite{Campbell:2022qmc}. This includes the standardisation of
next-to-leading order matched event generation, either with the \MCatNLO \cite{Frixione:2002ik} or
\powheg \cite{Nason:2004rx} methods and merging techniques at LO and NLO accuracy. Both have been
developed in full generality, but have received less attention in the context of
DIS over the past years. Further, jet-substructure techniques
\cite{Marzani:2019hun} receive significant attention at the LHC \cite{Andersen:2024czj},
and can find applications in a DIS context as well. One example is the soft-drop
grooming method \cite{Larkoski:2014wba}, that has been
generalised and applied also to jets at lepton colliders \cite{Baron:2018nfz,
  Chen:2021uws} and event shapes at both lepton colliders \cite{Baron:2018nfz,
  Marzani:2019evv}, in Higgs boson decays \cite{Gehrmann-DeRidder:2024avt} and
at hadron colliders \cite{Baron:2020xoi}, and, recently, also to event shapes in 
DIS \cite{Makris:2021drz}.

In this context, the H1 collaboration recently made use of this version of 
soft-drop grooming, to reanalyse data and measure the groomed jet mass
as well as the groomed 1-jettiness \cite{H1:2024pvu}. Likewise, plain
1-jettiness was measured \cite{H1:2024aze}, which is equivalent to thrust but
formulated as a manifestly global event shape. We here report on the
semi-analytical predictions used in these measurements, that were originally
derived in \cite{Knobbe:2023ehi}.

DIS measurements further offer a wealth of data valuable for tuning non-perturbative
model parameters in Monte Carlo event generators. These data are largely complementary 
to both lepton colliders, without initial-state hadrons, and hadron colliders, where
often effects from the underlying event dominate. This was exploited in \cite{Knobbe:2023ehi} to produce tunes of \sherpa's newly implemented hadronisation model \cite{Chahal:2022rid}, 
including replica tunes in the style of \cite{Knobbe:2023njd} for uncertainty estimates.

We will here focus on predictions for the groomed-mass observable, that we
compile in the same framework as used in \cite{Knobbe:2023ehi} but have not
discussed in detail there. In Sec.~\ref{sec:dis} we introduce the technical
details of soft-drop grooming and the observable definitions. We proceed in
Sec.~\ref{sec:shdis} to discuss the Monte Carlo simulation of DIS events and in
Sec.~\ref{sec:resdis} our framework for resummation of event shapes in DIS. We
present our results at the example of the groomed mass in Sec.~\ref{sec:grmass} 
before concluding.

\section{Soft-drop groomed mass in DIS}\label{sec:dis}

Soft-drop grooming is a popular jet substructure technique used at the LHC in 
various contexts, see for example \cite{Larkoski:2015lea, Frye:2016aiz, Kang:2018vgn, Baron:2020xoi, Cal:2021fla, Caletti:2021oor, Reichelt:2021svh, Caletti:2022hnc}.
The general idea is to recluster a given object in a
collision event with a suitable jet algorithm to identify the branching history,
and then drop softer branches while proceeding reversely through this tree. The
DIS  specific version of \cite{Makris:2021drz} is based on the \Centauro jet
algorithm \cite{Arratia:2020ssx}. The distance measure between particles with
momenta $p_i, p_j$ in this algorithm is given by
\begin{equation}
  d_{ij} = (\Delta\bar{z}_{ij})^2 + 2\bar{z}_i\bar{z}_j(1-\cos\Delta\phi_{ij})\,,\;\; \text{with}\;\;\bar{z}_i = 2\sqrt{1+\frac{q\cdot p_i}{x_B P\cdot p_i}}\quad\text{and}\quad ~\Delta\bar{z}_{ij}=\bar{z}_{i}-\bar{z}_{j}\,,\label{eq:centauro_def_zb}
\end{equation}
where we are using the standard DIS kinematics with $P_\mu$ the proton momentum,
and the exchanged photon momentum given by the difference between the incoming and
outgoing electron momentum $q_\mu = k_\mu-k^\prime_\mu$. For future
reference we also define the usual variables
\begin{equation}
  Q^2 = -q^2~,\;\;x_B = \frac{Q^2}{2P\cdot q}~,\;\;\text{and}\;\;y=\frac{P\cdot q}{P\cdot k}\,.
\end{equation}
Two branches in the clustering history are compared using the measure
\begin{equation}\label{eq:soft_drop_inequ}
  \frac{\min[z_i,z_j]}{z_i+z_j} > \zcut~,\;\;\text{with}\;\;z_i = \frac{P\cdot p_i}{P\cdot q}\,,
\end{equation}
where $\zcut$ is an adjustable parameter of the grooming algorithm. If the
soft-drop condition is not satisfied the branch with smaller $z$ is dropped, and
the procedure is repeated with the other branch. The algorithm terminates if
either Eq.~\eqref{eq:soft_drop_inequ} is satisfied, or if there is only one particle left.

After applying the grooming algorithm, properties of the surviving final state
can be calculated as before. We focus on the mass of the groomed final state
\begin{equation}\label{eq:mass_def}
  \rho = \frac{\left(\sum_i p_i\right)^2}{Q_0^2}~,
\end{equation}
which has also been studied in \cite{Makris:2021drz}. The sum here extends over all
final-state hadrons that have not been dropped during the grooming procedure. To
be compatible with \cite{Makris:2021drz} and the measurement in
\cite{H1:2024pvu}, we normalise to the minimal $Q^2$ value considered in the
measurement, namely $Q_0^2=150~\text{GeV}^2$.

\section{\sherpa framework for DIS}\label{sec:shdis}
We derive hadron-level predictions for the DIS event shapes using a pre-release
version of \sherpa-3.0~\cite{sherpa:2019gpd,sherpa3.0.beta}. To analyse our
simulated event samples we employ the \rivet analysis package~\cite{Bierlich:2019rhm}.
For jet clustering we use the \Centauro plugin~\cite{Arratia:2020ssx} within the \fastjet
framework~\cite{Cacciari:2011ma}. 

We consider the massless single and dijet production
channels in neutral current DIS at next-to-leading order (NLO), and three- and 
four-jets at leading order (LO). In our simulation we consider $u,d,s$ 
quarks to be massless, and include single and dijet production at NLO, 
whereas processes involving massive $c,b$ quarks are added at LO \cite{Krauss:2016orf}. 
For all cases, three- and four-jet processes are 
included at LO. Different multiplicities are
consistently merged together according to the MEPS@NLO~\cite{Hoeche:2012yf} and
MEPS@LO~\cite{Hoeche:2009rj} truncated-shower prescriptions using the
Catani--Seymour dipole shower~\cite{Schumann:2007mg}. The DIS specific
adaptations to the merging formalism have originally been described
in~\cite{Carli:2010cg}. Tree-level matrix elements are provided by
\comix~\cite{Gleisberg:2008fv} and \amegic \cite{Krauss:2001iv}. As parton density 
functions we use
the NNLO PDF4LHC21\_40\_pdfas set~\cite{PDF4LHCWorkingGroup:2022cjn} with
$\alpha_S(M^2_Z)$=0.118 obtained from LHAPDF \cite{Buckley:2014ana}. Beyond the core process,
the arguments of the strong-coupling factors are determined by the clustering
algorithm~\cite{Hoeche:2009rj}, and we set the core scale as well as the
merging-scale parameter dynamically, thereby following Ref.~\cite{Carli:2010cg}. The events
get hadronised using \sherpa's new implementation of the cluster hadronisation
model~\cite{Chahal:2022rid}, tuned to LEP \cite{Knobbe:2023njd} and DIS \cite{Knobbe:2023ehi}
data, including replica tunes to estimate the uncertainty induced by non-perturbative corrections.

\section{Resummed predictions with \sherpa + \Caesar}\label{sec:resdis}

We derive predictions at NLL accuracy using the implementation of the \Caesar
formalism~\cite{Banfi:2004yd} available in the \sherpa
framework~\cite{Gerwick:2014gya,Baberuxki:2019ifp}. The formalism provides a
master formula, valid for recursive infrared and collinear (rIRC) safe
observables, for the cumulative cross section
integrating observable values up to $v = \exp(-L)$. For a 2-jet observable like
groomed jet mass in DIS it is written as follows:
\begin{equation}\label{eq:CAESAR}
  \begin{split}
    \Sigma_\mathrm{res}(v) &= \int d\mathcal{B}
    \frac{\mathop{d\sigma}}{\mathop{d\mathcal{B}}} \exp\left[-\sum_{l}
      R_l^\mathcal{B}(L)\right]\mathcal{P}^{\mathcal{B}}(L)\mathcal{F}^\mathcal{B}(L)\mathcal{H}(\mathcal{B})\,,
  \end{split}
\end{equation}
where $\frac{\mathop{d\sigma}}{\mathop{d\mathcal{B}}}$ is the
fully differential Born cross section and $\mathcal{H}$ implements the kinematic
cuts applied to the Born phase space $\mathcal{B}$. Since we are
dealing with an additive observable, the multiple emission function $\mathcal{F}$
is simply given by  $\mathcal{F}(L) = e^{-\gamma_E
  R^\prime}/\Gamma(1+R^\prime)$, with $R^\prime(L)=\partial R/\partial L$ and
$R(L)=\sum_{l} R_l(L)$.  The collinear radiators $R_l$ for the hard legs
$l$ were computed in~\cite{Banfi:2004yd}. We match our resummed calculation in
the multiplicative matching scheme along the lines of \cite{Baberuxki:2019ifp}.
The extensions made in \cite{Baron:2020xoi} to accommodate the phase-space
constraints implied by soft-drop grooming with general parameters $\zcut$ and
$\beta$, and that have been used to describe groomed jet substructure in
\cite{Caletti:2021oor, Caletti:2021ysv, Reichelt:2021svh}, are directly
applicable here. For the detailed chain of arguments see also
\cite{Knobbe:2023ehi}.

We use the functionality of \sherpa as a matrix element generator and fixed order Monte Carlo
to produce $\mathcal{O}(\alphaS^2)$ accurate differential predictions for normalised event shape distributions.
This can be achieved by considering an NLO calculation of the 2-jet production process with a cut requiring
a minimal value of the considered observable. This cut can be chosen to be smaller than any bin resolved by
the experiment (or otherwise of interest). The missing contributions completely drop out of normalised distributions.
Here however, we want to include the total cross-section prediction. We do this by computing the NNLO accurate cross-section using the projection-to-Born method \cite{Cacciari:2015jma} that has been automated in \sherpa for DIS \cite{Hoche:2018gti} and the Drell-Yan process \cite{Hoche:2014uhw}. The cross section differential in $y$ and $Q^2$ is sufficient to fix the missing contributions to the differential cross section. Note that while at fixed order that contribution is confined to the bin including a value of the event shape of 0, this is not necessarily true in our multiplicative matching scheme.

While soft-drop grooming has been shown to reduce the impact of non-perturbative
corrections in various circumstances, for example in \cite{Frye:2016aiz, Baron:2018nfz, Marzani:2019evv,
  Baron:2020xoi, Caletti:2021oor, Reichelt:2021svh, dEnterria:2022hzv, Chien:2024uax}, it is typically still
necessary to account for a remaining small hadronisation contribution. We
here adopt the approach of \cite{Reichelt:2021svh} to extract transfer matrices
from \sherpa Monte Carlo simulations. This approach has been shown to be
superior to bin-wise ratios between hadron and parton level Monte Carlo, see
Refs. \cite{Reichelt:2021svh,Chien:2024uax}, and has been connected to the shape
function approach \cite{Korchemsky:1999kt} in \cite{Chien:2024uax}.

\section{Results for groomed jet mass}\label{sec:grmass}

We derive predictions for the groomed invariant mass $\rho$ in the phase-space region 
$0.2<y<0.7$ and $\unit[150]{GeV^2}<Q^2<\unit[20000]{GeV^2}$.
To estimate perturbative uncertainties, we consider 7-point variations of the 
factorisation and renormalisation scales in the matrix element and
the parton shower that get evaluated on-the-fly \cite{Bothmann:2016nao}.
The resummation scale we keep fixed. We estimate the impact of sub-leading
logarithms in the resummation by varying $x_L$ in the form of the logarithm according to
\beq
  L\to\ln\left(\frac{x_L}{v}-x_L+1\right)\xrightarrow{\substack{v \to 0}}\ln\left(\frac{x_L}{v}\right),
\eeq
assuming $x_L\in \{0.5,1,2\}$, leaving the distribution at the kinematic endpoint 
unchanged. Furthermore, we consider non-perturbative uncertainties through a set of
replica tunes, for that we extract individual transfer matrices. The final systematic uncertainty estimate is derived by forming an envelope of all variations for the 
MEPS@NLO distribution and of all combinations of the scale, $x_L$ and transfer matrix 
variations for the resummed distribution. 

\begin{figure}
  \centering
  \includegraphics[width=0.32\textwidth]{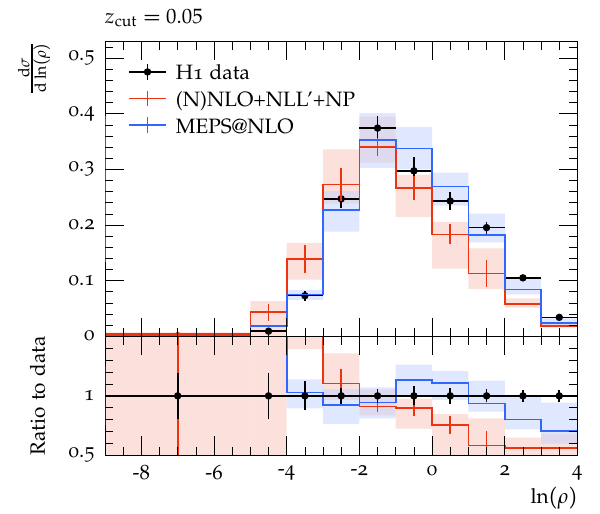}
  \includegraphics[width=0.32\textwidth]{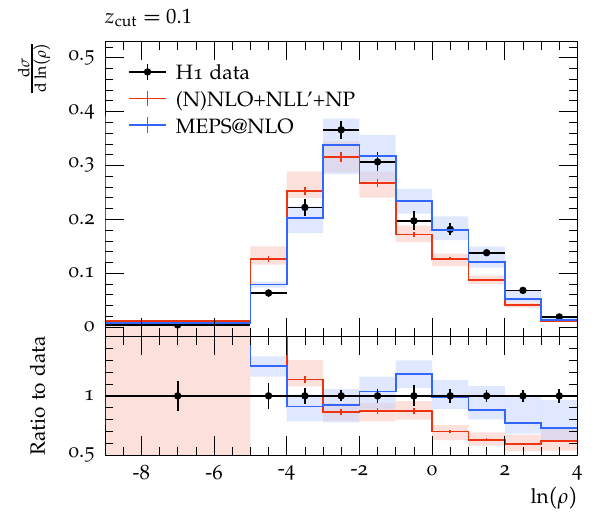}
  \includegraphics[width=0.32\textwidth]{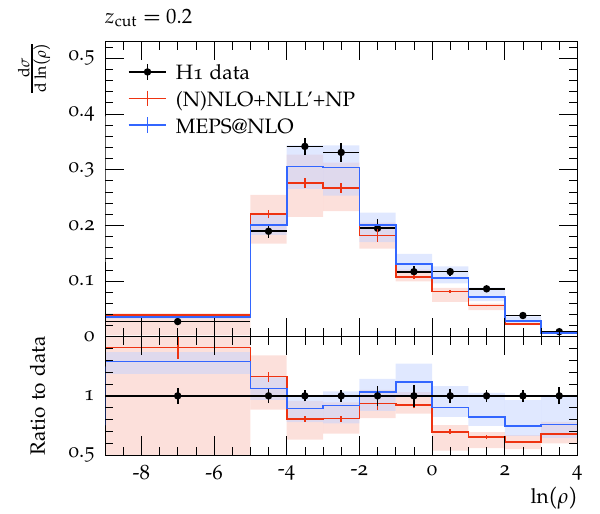}
  \caption{Differential cross section of the natural logarithm of the groomed invariant mass $\ln(\rho)$ in DIS at $\sqrt{s}=\unit[319]{GeV}$ for $\zcut\in\{0.05,0.1,0.2\}$ (left to right panel) and $\beta=0$. The phase space is restricted to $0.2<y<0.7$ and $\unit[150]{GeV^2}<Q^2<\unit[20000]{GeV^2}$. The bars on the data points illustrate the combined statistical and systematic uncertainty. The systematic uncertainties of the theory predictions are shown by the colored envelopes, while the error bars depict the statistical uncertainties.}
  \label{fig:grmass}
\end{figure}

Fig.~\ref{fig:grmass} compares hadron-level MEPS@NLO simulations from \sherpa and  (N)NLO+NLL'+NP predictions to H1 data from \cite{H1:2024pvu} for the differential cross section of the groomed invariant mass for soft-drop parameter 
$\zcut\in\{0.05,0.1,0.2\}$. \sherpa provides a good description of the data distributions, independent of the value of $\zcut$, largely consistent within uncertainties. Only the first two bins show a larger deviation between prediction and data, with \sherpa overestimating the data. The resummed results matched to 
\NLO and including non-perturbative corrections predict a lower mass than observed in data. However, the agreement improves for stronger grooming  where non-perturbative corrections are reduced. 

\begin{figure}
  \centering
  \includegraphics[width=\textwidth]{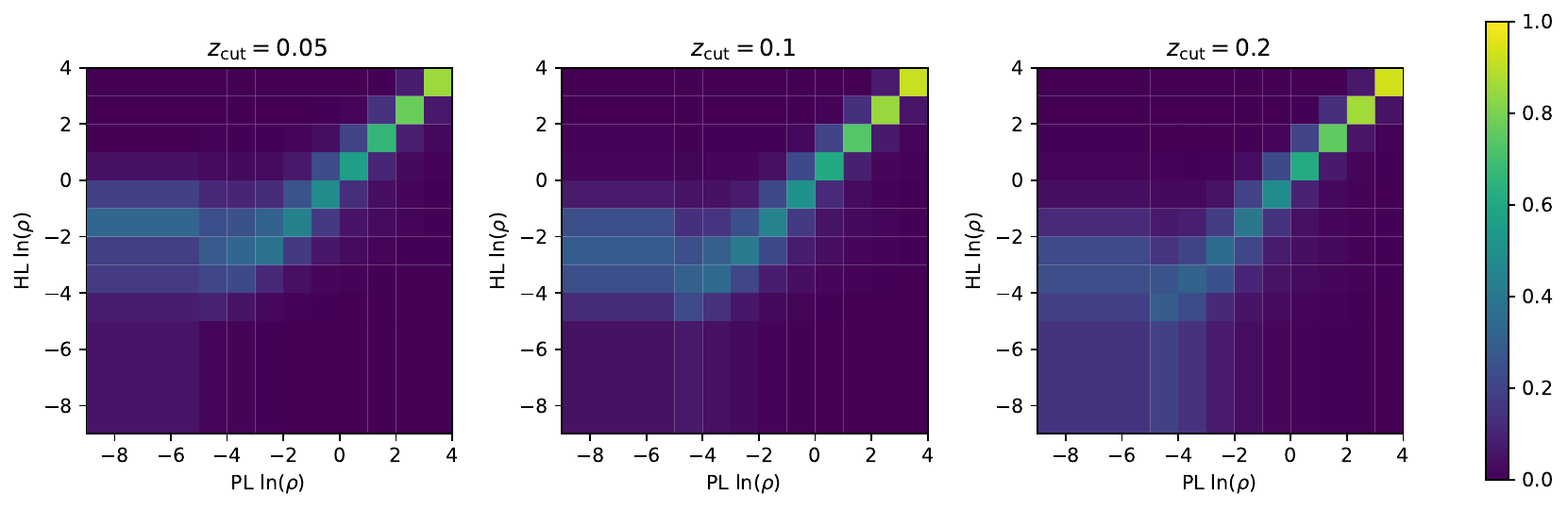}
  \caption{Transfer matrices (corresponding to the default tuning parameter set) used to account for hadronisation corrections to the resummed predictions for the groomed invariant mass for $\zcut\in\{0.05,0.1,0.2\}$ (left to right). Each matrix entry $T_{ij}$ describes the probability for an event in parton level (PL) bin $j$ to migrate into the hadron level (HL) bin $i$.}
  \label{fig:tmatrices}
\end{figure}

This is supported when considering the transfer matrices shown in Fig.~\ref{fig:tmatrices} that are used to account for hadronisation corrections. 
With higher $\zcut$ values the parton-to-hadron level migration matrices become more
centered around the diagonal. In particular for $\zcut=0.05$ for PL $\ln(\rho)<-2$ the shifts in the observable towards larger values at HL are quite significant. However, in the region where the largest deviations from data are observed, \emph{i.e.}\ for large values of $\ln(\rho)$, non-perturbative corrections are 
indeed rather mild. In turn, the discrepancy between data and resummation at large $\ln(\rho)$ presumably originate from the lack of higher-order hard-emission contributions. 

\section{Conclusion}

We have studied the groomed jet-mass observable in DIS as an example of a groomed event shape and supplementing the results of \cite{Knobbe:2023ehi} for 1-jettiness. Our predictions are compared to
data from the H1 experiment \cite{H1:2024pvu} where recently also the 1-jettiness observable was studied \cite{H1:2024aze}. An interesting complementary measurement in this context is the distribution of events with an empty 
current hemisphere \cite{H1:2024nde}. We achieve NNLO accuracy for the DIS cross section in the event-selection 
phase space, corresponding to NLO accuracy for the event shapes we consider. We match this to an NLL accurate
calculation obtained within the \Caesar formalism, to achieve overall (N)\NLOpNLLp accuracy. Hadronisation corrections 
are included in the transfer-matrix approach based on Monte Carlo simulations with the \sherpa event generator. We also
showcase those predictions at MEPS@NLO accuracy, enabling the critically needed comparison between parton showers and analytic resummation \cite{Hoche:2017kst}. Both the (N)\NLOpNLLp result and the \sherpa sample give a fair
description of the groomed mass. Notable deficiencies are observable in the small-mass limit, susceptible both to all-orders perturbative as well as non-perturbative corrections. This is consistent
with the observations in \cite{H1:2024pvu,H1:2024aze}, where excellent agreement for the corresponding predictions
for 1-jettiness was found, however not as closely examining the soft limit on a logarithmic scale like the groomed
jet-mass measurement does. Note, the predictions for the groomed-mass observable presented here are also discussed in \cite{H1:2024pvu}, where a more extensive comparison to data differential in $y$ and $Q^2$ bins can be found.

\acknowledgments
We would like to thank the H1 collaboration and in particular Daniel Britzger, Henry Klest, Johannes Hessler and 
Vinicius Mikuni for fruitful discussions on this and related projects.
We are indebted to Stefan H\"oche for assistance with the NNLO corrections and to Frank Krauss for
help with \sherpa's new beam fragmentation model.
MK, SS and LS acknowledge support from BMBF (05H21MGCAB) and funding by the Deutsche Forschungsgemeinschaft
(DFG, German Research Foundation) - project number 456104544 and 510810461. 
DR is supported by the STFC IPPP grant (ST/T001011/1).

\bibliographystyle{JHEP}
\bibliography{references}

\providecommand{\href}[2]{#2}\begingroup\raggedright\begin{thebibliography}{10}

\bibitem{Campbell:2022qmc}
J.M.~Campbell et~al., \emph{{Event Generators for High-Energy Physics
  Experiments}},
  \href{https://doi.org/10.21468/SciPostPhys.16.5.130}{\emph{SciPost Phys.}
  {\bfseries 16} (2024) 130}
  [\href{https://arxiv.org/abs/2203.11110}{{\ttfamily 2203.11110}}].

\bibitem{Frixione:2002ik}
S.~Frixione and B.R.~Webber, \emph{{Matching NLO QCD computations and parton
  shower simulations}},
  \href{https://doi.org/10.1088/1126-6708/2002/06/029}{\emph{JHEP} {\bfseries
  06} (2002) 029} [\href{https://arxiv.org/abs/hep-ph/0204244}{{\ttfamily
  hep-ph/0204244}}].

\bibitem{Nason:2004rx}
P.~Nason, \emph{{A New method for combining NLO QCD with shower Monte Carlo
  algorithms}},
  \href{https://doi.org/10.1088/1126-6708/2004/11/040}{\emph{JHEP} {\bfseries
  11} (2004) 040} [\href{https://arxiv.org/abs/hep-ph/0409146}{{\ttfamily
  hep-ph/0409146}}].

\bibitem{Marzani:2019hun}
S.~Marzani, G.~Soyez and M.~Spannowsky, \emph{{Looking inside jets: an
  introduction to jet substructure and boosted-object phenomenology}},
  vol.~958, Springer (2019),
  \href{https://doi.org/10.1007/978-3-030-15709-8}{10.1007/978-3-030-15709-8},
  [\href{https://arxiv.org/abs/1901.10342}{{\ttfamily 1901.10342}}].

\bibitem{Andersen:2024czj}
J.~Andersen et~al., \emph{{Les Houches 2023: Physics at TeV Colliders: Standard
  Model Working Group Report}},
  \href{https://arxiv.org/abs/2406.00708}{{\ttfamily 2406.00708}}.

\bibitem{Larkoski:2014wba}
A.J.~Larkoski, S.~Marzani, G.~Soyez and J.~Thaler, \emph{{Soft Drop}},
  \href{https://doi.org/10.1007/JHEP05(2014)146}{\emph{JHEP} {\bfseries 05}
  (2014) 146} [\href{https://arxiv.org/abs/1402.2657}{{\ttfamily 1402.2657}}].

\bibitem{Baron:2018nfz}
J.~Baron, S.~Marzani and V.~Theeuwes, \emph{{Soft-Drop Thrust}},
  \href{https://doi.org/10.1007/JHEP08(2018)105}{\emph{JHEP} {\bfseries 08}
  (2018) 105} [\href{https://arxiv.org/abs/1803.04719}{{\ttfamily
  1803.04719}}].

\bibitem{Chen:2021uws}
Y.~Chen et~al., \emph{{Jet energy spectrum and substructure in e$^{+}$e$^{-}$
  collisions at 91.2 GeV with ALEPH Archived Data}},
  \href{https://doi.org/10.1007/JHEP06(2022)008}{\emph{JHEP} {\bfseries 06}
  (2022) 008} [\href{https://arxiv.org/abs/2111.09914}{{\ttfamily
  2111.09914}}].

\bibitem{Marzani:2019evv}
S.~Marzani, D.~Reichelt, S.~Schumann, G.~Soyez and V.~Theeuwes, \emph{{Fitting
  the Strong Coupling Constant with Soft-Drop Thrust}},
  \href{https://doi.org/10.1007/JHEP11(2019)179}{\emph{JHEP} {\bfseries 11}
  (2019) 179} [\href{https://arxiv.org/abs/1906.10504}{{\ttfamily
  1906.10504}}].

\bibitem{Gehrmann-DeRidder:2024avt}
A.~Gehrmann-De~Ridder, C.T.~Preuss, D.~Reichelt and S.~Schumann,
  \emph{{NLO+NLL' accurate predictions for three-jet event shapes in hadronic
  Higgs decays}},  \href{https://arxiv.org/abs/2403.06929}{{\ttfamily
  2403.06929}}.

\bibitem{Baron:2020xoi}
J.~Baron, D.~Reichelt, S.~Schumann, N.~Schwanemann and V.~Theeuwes,
  \emph{{Soft-drop grooming for hadronic event shapes}},
  \href{https://doi.org/10.1007/JHEP07(2021)142}{\emph{JHEP} {\bfseries 07}
  (2021) 142} [\href{https://arxiv.org/abs/2012.09574}{{\ttfamily
  2012.09574}}].

\bibitem{Makris:2021drz}
Y.~Makris, \emph{{Revisiting the role of grooming in DIS}},
  \href{https://doi.org/10.1103/PhysRevD.103.054005}{\emph{Phys. Rev. D}
  {\bfseries 103} (2021) 054005}
  [\href{https://arxiv.org/abs/2101.02708}{{\ttfamily 2101.02708}}].

\bibitem{H1:2024pvu}
{\scshape H1} collaboration, \emph{{Measurement of groomed event shape
  observables in deep-inelastic electron-proton scattering at HERA}},
  \href{https://arxiv.org/abs/2403.10134}{{\ttfamily 2403.10134}}.

\bibitem{H1:2024aze}
{\scshape H1} collaboration, \emph{{Measurement of the 1-jettiness event shape
  observable in deep-inelastic electron-proton scattering at HERA}},
  \href{https://arxiv.org/abs/2403.10109}{{\ttfamily 2403.10109}}.

\bibitem{Knobbe:2023ehi}
M.~Knobbe, D.~Reichelt and S.~Schumann, \emph{{(N)NLO+NLL\textquoteright{}
  accurate predictions for plain and groomed 1-jettiness in neutral current
  DIS}}, \href{https://doi.org/10.1007/JHEP09(2023)194}{\emph{JHEP} {\bfseries
  09} (2023) 194} [\href{https://arxiv.org/abs/2306.17736}{{\ttfamily
  2306.17736}}].

\bibitem{Chahal:2022rid}
G.S.~Chahal and F.~Krauss, \emph{{Cluster Hadronisation in Sherpa}},
  \href{https://doi.org/10.21468/SciPostPhys.13.2.019}{\emph{SciPost Phys.}
  {\bfseries 13} (2022) 019}
  [\href{https://arxiv.org/abs/2203.11385}{{\ttfamily 2203.11385}}].

\bibitem{Knobbe:2023njd}
M.~Knobbe, F.~Krauss, D.~Reichelt and S.~Schumann, \emph{{Measuring hadronic
  Higgs boson branching ratios at future lepton colliders}},
  \href{https://doi.org/10.1140/epjc/s10052-024-12430-4}{\emph{Eur. Phys. J. C}
  {\bfseries 84} (2024) 83} [\href{https://arxiv.org/abs/2306.03682}{{\ttfamily
  2306.03682}}].

\bibitem{Larkoski:2015lea}
A.J.~Larkoski, S.~Marzani and J.~Thaler, \emph{{Sudakov Safety in Perturbative
  QCD}}, \href{https://doi.org/10.1103/PhysRevD.91.111501}{\emph{Phys. Rev. D}
  {\bfseries 91} (2015) 111501}
  [\href{https://arxiv.org/abs/1502.01719}{{\ttfamily 1502.01719}}].

\bibitem{Frye:2016aiz}
C.~Frye, A.J.~Larkoski, M.D.~Schwartz and K.~Yan, \emph{{Factorization for
  groomed jet substructure beyond the next-to-leading logarithm}},
  \href{https://doi.org/10.1007/JHEP07(2016)064}{\emph{JHEP} {\bfseries 07}
  (2016) 064} [\href{https://arxiv.org/abs/1603.09338}{{\ttfamily
  1603.09338}}].

\bibitem{Kang:2018vgn}
Z.-B.~Kang, K.~Lee, X.~Liu and F.~Ringer, \emph{{Soft drop groomed jet
  angularities at the LHC}},
  \href{https://doi.org/10.1016/j.physletb.2019.04.018}{\emph{Phys. Lett. B}
  {\bfseries 793} (2019) 41}
  [\href{https://arxiv.org/abs/1811.06983}{{\ttfamily 1811.06983}}].

\bibitem{Cal:2021fla}
P.~Cal, K.~Lee, F.~Ringer and W.J.~Waalewijn, \emph{{The soft drop momentum
  sharing fraction zg beyond leading-logarithmic accuracy}},
  \href{https://doi.org/10.1016/j.physletb.2022.137390}{\emph{Phys. Lett. B}
  {\bfseries 833} (2022) 137390}
  [\href{https://arxiv.org/abs/2106.04589}{{\ttfamily 2106.04589}}].

\bibitem{Caletti:2021oor}
S.~Caletti, O.~Fedkevych, S.~Marzani, D.~Reichelt, S.~Schumann, G.~Soyez
  et~al., \emph{{Jet angularities in Z+jet production at the LHC}},
  \href{https://doi.org/10.1007/JHEP07(2021)076}{\emph{JHEP} {\bfseries 07}
  (2021) 076} [\href{https://arxiv.org/abs/2104.06920}{{\ttfamily
  2104.06920}}].

\bibitem{Reichelt:2021svh}
D.~Reichelt, S.~Caletti, O.~Fedkevych, S.~Marzani, S.~Schumann and G.~Soyez,
  \emph{{Phenomenology of jet angularities at the LHC}},
  \href{https://doi.org/10.1007/JHEP03(2022)131}{\emph{JHEP} {\bfseries 03}
  (2022) 131} [\href{https://arxiv.org/abs/2112.09545}{{\ttfamily
  2112.09545}}].

\bibitem{Caletti:2022hnc}
S.~Caletti, A.J.~Larkoski, S.~Marzani and D.~Reichelt, \emph{{Practical jet
  flavour through NNLO}},
  \href{https://doi.org/10.1140/epjc/s10052-022-10568-7}{\emph{Eur. Phys. J. C}
  {\bfseries 82} (2022) 632}
  [\href{https://arxiv.org/abs/2205.01109}{{\ttfamily 2205.01109}}].

\bibitem{Arratia:2020ssx}
M.~Arratia, Y.~Makris, D.~Neill, F.~Ringer and N.~Sato, \emph{{Asymmetric jet
  clustering in deep-inelastic scattering}},
  \href{https://doi.org/10.1103/PhysRevD.104.034005}{\emph{Phys. Rev. D}
  {\bfseries 104} (2021) 034005}
  [\href{https://arxiv.org/abs/2006.10751}{{\ttfamily 2006.10751}}].

\bibitem{sherpa:2019gpd}
{\scshape Sherpa} collaboration, \emph{{Event Generation with Sherpa 2.2}},
  \href{https://doi.org/10.21468/SciPostPhys.7.3.034}{\emph{SciPost Phys.}
  {\bfseries 7} (2019) 034} [\href{https://arxiv.org/abs/1905.09127}{{\ttfamily
  1905.09127}}].

\bibitem{sherpa3.0.beta}
``\normalfont{The \sherpa-3.0.beta code can be obtained from:}
  \url{https://sherpa-team.gitlab.io/changelog.html}.''

\bibitem{Bierlich:2019rhm}
C.~Bierlich et~al., \emph{{Robust Independent Validation of Experiment and
  Theory: Rivet version 3}},
  \href{https://doi.org/10.21468/SciPostPhys.8.2.026}{\emph{SciPost Phys.}
  {\bfseries 8} (2020) 026} [\href{https://arxiv.org/abs/1912.05451}{{\ttfamily
  1912.05451}}].

\bibitem{Cacciari:2011ma}
M.~Cacciari, G.P.~Salam and G.~Soyez, \emph{{FastJet User Manual}},
  \href{https://doi.org/10.1140/epjc/s10052-012-1896-2}{\emph{Eur. Phys. J.}
  {\bfseries C72} (2012) 1896}
  [\href{https://arxiv.org/abs/1111.6097}{{\ttfamily 1111.6097}}].

\bibitem{Krauss:2016orf}
F.~Krauss, D.~Napoletano and S.~Schumann, \emph{{Simulating $b$-associated
  production of $Z$ and Higgs bosons with the SHERPA event generator}},
  \href{https://doi.org/10.1103/PhysRevD.95.036012}{\emph{Phys. Rev. D}
  {\bfseries 95} (2017) 036012}
  [\href{https://arxiv.org/abs/1612.04640}{{\ttfamily 1612.04640}}].

\bibitem{Hoeche:2012yf}
S.~H{\"o}che, F.~Krauss, M.~Sch{\"o}nherr and F.~Siegert, \emph{{QCD matrix
  elements + parton showers: The NLO case}},
  \href{https://doi.org/10.1007/JHEP04(2013)027}{\emph{JHEP} {\bfseries 04}
  (2013) 027} [\href{https://arxiv.org/abs/1207.5030}{{\ttfamily 1207.5030}}].

\bibitem{Hoeche:2009rj}
S.~H{\"o}che, F.~Krauss, S.~Schumann and F.~Siegert, \emph{{QCD matrix elements
  and truncated showers}},
  \href{https://doi.org/10.1088/1126-6708/2009/05/053}{\emph{JHEP} {\bfseries
  05} (2009) 053} [\href{https://arxiv.org/abs/0903.1219}{{\ttfamily
  0903.1219}}].

\bibitem{Schumann:2007mg}
S.~Schumann and F.~Krauss, \emph{{A Parton shower algorithm based on
  Catani-Seymour dipole factorisation}},
  \href{https://doi.org/10.1088/1126-6708/2008/03/038}{\emph{JHEP} {\bfseries
  03} (2008) 038} [\href{https://arxiv.org/abs/0709.1027}{{\ttfamily
  0709.1027}}].

\bibitem{Carli:2010cg}
T.~Carli, T.~Gehrmann and S.~H{\"o}che, \emph{{Hadronic final states in
  deep-inelastic scattering with Sherpa}},
  \href{https://doi.org/10.1140/epjc/s10052-010-1261-2}{\emph{Eur. Phys. J. C}
  {\bfseries 67} (2010) 73} [\href{https://arxiv.org/abs/0912.3715}{{\ttfamily
  0912.3715}}].

\bibitem{Gleisberg:2008fv}
T.~Gleisberg and S.~H{\"o}che, \emph{{Comix, a new matrix element generator}},
  \href{https://doi.org/10.1088/1126-6708/2008/12/039}{\emph{JHEP} {\bfseries
  12} (2008) 039} [\href{https://arxiv.org/abs/0808.3674}{{\ttfamily
  0808.3674}}].

\bibitem{Krauss:2001iv}
F.~Krauss, R.~Kuhn and G.~Soff, \emph{{AMEGIC++ 1.0: A Matrix element generator
  in C++}}, \href{https://doi.org/10.1088/1126-6708/2002/02/044}{\emph{JHEP}
  {\bfseries 02} (2002) 044}
  [\href{https://arxiv.org/abs/hep-ph/0109036}{{\ttfamily hep-ph/0109036}}].

\bibitem{PDF4LHCWorkingGroup:2022cjn}
{\scshape PDF4LHC Working Group} collaboration, \emph{{The PDF4LHC21
  combination of global PDF fits for the LHC Run III}},
  \href{https://doi.org/10.1088/1361-6471/ac7216}{\emph{J. Phys. G} {\bfseries
  49} (2022) 080501} [\href{https://arxiv.org/abs/2203.05506}{{\ttfamily
  2203.05506}}].

\bibitem{Buckley:2014ana}
A.~Buckley, J.~Ferrando, S.~Lloyd, K.~Nordstr\"om, B.~Page, M.~R\"ufenacht
  et~al., \emph{{LHAPDF6: parton density access in the LHC precision era}},
  \href{https://doi.org/10.1140/epjc/s10052-015-3318-8}{\emph{Eur. Phys. J. C}
  {\bfseries 75} (2015) 132} [\href{https://arxiv.org/abs/1412.7420}{{\ttfamily
  1412.7420}}].

\bibitem{Banfi:2004yd}
A.~Banfi, G.P.~Salam and G.~Zanderighi, \emph{{Principles of general
  final-state resummation and automated implementation}},
  \href{https://doi.org/10.1088/1126-6708/2005/03/073}{\emph{JHEP} {\bfseries
  03} (2005) 073} [\href{https://arxiv.org/abs/hep-ph/0407286}{{\ttfamily
  hep-ph/0407286}}].

\bibitem{Gerwick:2014gya}
E.~Gerwick, S.~H{\"o}che, S.~Marzani and S.~Schumann, \emph{{Soft evolution of
  multi-jet final states}},
  \href{https://doi.org/10.1007/JHEP02(2015)106}{\emph{JHEP} {\bfseries 02}
  (2015) 106} [\href{https://arxiv.org/abs/1411.7325}{{\ttfamily 1411.7325}}].

\bibitem{Baberuxki:2019ifp}
N.~Baberuxki, C.T.~Preuss, D.~Reichelt and S.~Schumann, \emph{{Resummed
  predictions for jet-resolution scales in multijet production in
  e$^{+}$e$^{-}$ annihilation}},
  \href{https://doi.org/10.1007/JHEP04(2020)112}{\emph{JHEP} {\bfseries 04}
  (2020) 112} [\href{https://arxiv.org/abs/1912.09396}{{\ttfamily
  1912.09396}}].

\bibitem{Caletti:2021ysv}
S.~Caletti, O.~Fedkevych, S.~Marzani and D.~Reichelt, \emph{{Tagging the
  initial-state gluon}},
  \href{https://doi.org/10.1140/epjc/s10052-021-09648-x}{\emph{Eur. Phys. J. C}
  {\bfseries 81} (2021) 844}
  [\href{https://arxiv.org/abs/2108.10024}{{\ttfamily 2108.10024}}].

\bibitem{Cacciari:2015jma}
M.~Cacciari, F.A.~Dreyer, A.~Karlberg, G.P.~Salam and G.~Zanderighi,
  \emph{{Fully Differential Vector-Boson-Fusion Higgs Production at
  Next-to-Next-to-Leading Order}},
  \href{https://doi.org/10.1103/PhysRevLett.115.082002}{\emph{Phys. Rev. Lett.}
  {\bfseries 115} (2015) 082002}
  [\href{https://arxiv.org/abs/1506.02660}{{\ttfamily 1506.02660}}].

\bibitem{Hoche:2018gti}
S.~H\"oche, S.~Kuttimalai and Y.~Li, \emph{{Hadronic Final States in DIS at
  NNLO QCD with Parton Showers}},
  \href{https://doi.org/10.1103/PhysRevD.98.114013}{\emph{Phys. Rev. D}
  {\bfseries 98} (2018) 114013}
  [\href{https://arxiv.org/abs/1809.04192}{{\ttfamily 1809.04192}}].

\bibitem{Hoche:2014uhw}
S.~H\"oche, Y.~Li and S.~Prestel, \emph{{Drell-Yan lepton pair production at
  NNLO QCD with parton showers}},
  \href{https://doi.org/10.1103/PhysRevD.91.074015}{\emph{Phys. Rev. D}
  {\bfseries 91} (2015) 074015}
  [\href{https://arxiv.org/abs/1405.3607}{{\ttfamily 1405.3607}}].

\bibitem{dEnterria:2022hzv}
D.~d'Enterria et~al., \emph{{The strong coupling constant: State of the art and
  the decade ahead}},  \href{https://arxiv.org/abs/2203.08271}{{\ttfamily
  2203.08271}}.

\bibitem{Chien:2024uax}
Y.-T.~Chien, O.~Fedkevych, D.~Reichelt and S.~Schumann, \emph{{Jet angularities
  in dijet production in proton-proton and heavy-ion collisions at RHIC}},
  \href{https://arxiv.org/abs/2404.04168}{{\ttfamily 2404.04168}}.

\bibitem{Korchemsky:1999kt}
G.P.~Korchemsky and G.F.~Sterman, \emph{{Power corrections to event shapes and
  factorization}},
  \href{https://doi.org/10.1016/S0550-3213(99)00308-9}{\emph{Nucl. Phys. B}
  {\bfseries 555} (1999) 335}
  [\href{https://arxiv.org/abs/hep-ph/9902341}{{\ttfamily hep-ph/9902341}}].

\bibitem{Bothmann:2016nao}
E.~Bothmann, M.~Sch\"onherr and S.~Schumann, \emph{{Reweighting QCD
  matrix-element and parton-shower calculations}},
  \href{https://doi.org/10.1140/epjc/s10052-016-4430-0}{\emph{Eur. Phys. J. C}
  {\bfseries 76} (2016) 590}
  [\href{https://arxiv.org/abs/1606.08753}{{\ttfamily 1606.08753}}].

\bibitem{H1:2024nde}
{\scshape H1} collaboration, \emph{{Observation and differential cross section
  measurement of neutral current DIS events with an empty hemisphere in the
  Breit frame}},  \href{https://arxiv.org/abs/2403.08982}{{\ttfamily
  2403.08982}}.

\bibitem{Hoche:2017kst}
S.~H\"oche, D.~Reichelt and F.~Siegert, \emph{{Momentum conservation and
  unitarity in parton showers and NLL resummation}},
  \href{https://doi.org/10.1007/JHEP01(2018)118}{\emph{JHEP} {\bfseries 01}
  (2018) 118} [\href{https://arxiv.org/abs/1711.03497}{{\ttfamily
  1711.03497}}].

\end{thebibliography}\endgroup

\end{document}